\documentclass[12pt]{iopart}
\usepackage{iopams}
\usepackage{setstack}
\usepackage{psfig}

\newcommand{\cP}{\ensuremath{\mathcal{P}}}
\newcommand{\cT}{\ensuremath{\mathcal{T}}}
\newcommand{\half}{\mbox{$\textstyle{\frac{1}{2}}$}}

\begin{document}
\rightline{preprint LA-UR-06-6952}
\title[$\cP\cT$-Symmetric Extension of the Korteweg-de Vries Equation]{$\cP
\cT$-Symmetric Extension of the Korteweg-de Vries Equation}

\author[Bender, Brody, Chen, and Furlan]{Carl~M~Bender$^*$\footnote{Permanent
address: Department of Physics, Washington University, St. Louis MO 63130, USA},
Dorje~C~Brody$^\dag$, Junhua Chen$^\ddag$, and Elisabetta Furlan$^{\dag}$}

\address{${}^*$Center for Nonlinear Studies, Los Alamos National Laboratory,
Los Alamos, NM 87545, USA}

\address{${}^\dag$Blackett Laboratory, Imperial College, London SW7 2BZ, UK}

\address{${}^\ddag$Department of Physics, Washington University, St. Louis MO
63130, USA}

\begin{abstract}
The Korteweg-de Vries equation $u_t+uu_x+u_{xxx}=0$ is $\cP\cT$ symmetric
(invariant under space-time reflection). Therefore, it can be generalized and
extended into the complex domain in such a way as to preserve the $\cP\cT$
symmetry. The result is the family of complex nonlinear wave equations $u_t-iu(i
u_x)^\epsilon+u_{xxx}=0$, where $\epsilon$ is real. The features of these
equations are discussed. Special attention is given to the $\epsilon=3$
equation, for which conservation laws are derived and solitary waves are
investigated.
\end{abstract}
\pacs{03.65.Ge, 02.60.Lj, 11.30.Er, 03.50.-z}
\submitto{\JPA}

Many papers have been written on theories described by non-Hermitian $\cP
\cT$-symmetric quantum-mechanical Hamiltonians. To construct such theories one
begins with a Hamiltonian that is both Hermitian and $\cP\cT$ symmetric, such
as the harmonic oscillator $H=p^2+x^2$. One then introduces a real parameter
$\epsilon$ to extend the Hamiltonian into the complex domain in such a way as to
preserve the $\cP\cT$ symmetry:
\begin{equation}
H=p^2+x^2(ix)^\epsilon.
\label{e1}
\end{equation}
The result is a family of complex non-Hermitian Hamiltonians that for positive
$\epsilon$ maintain many of the properties of the harmonic oscillator
Hamiltonian; namely, that the eigenvalues remain real, positive, and discrete 
\cite{r1,r2,r3}. The properties of classical $\cP\cT$-symmetric Hamiltonians
have also been examined \cite{r4,r5,r6,r7}. However, there are to date no
published studies of $\cP\cT$-symmetric classical wave equations.

The starting point of this paper is the heretofore unnoticed property that the
Korteweg-de Vries (KdV) equation,
\begin{equation}
u_t+uu_x+u_{xxx}=0,
\label{e2}
\end{equation}
is $\cP\cT$ symmetric. To demonstrate this, we define parity reflection $\cP$ by
$x\to-x$, and since $u=u(x,t)$ is a velocity, the sign of $u$ also changes under
$\cP$: $u\to-u$. We define time reversal $\cT$ by $t\to-t$, and again, since $u$
is a velocity, the sign of $u$ also changes under $\cT$: $u\to-u$. Following the
quantum-mechanical formalism, we also require that $i\to-i$ under time reversal.
It is clear that the KdV equation is not symmetric under $\cP$ or $\cT$
separately, but it {\it is} symmetric under combined $\cP\cT$ reflection. The
KdV equation is a special case of the Camassa-Holm equation \cite{r8}, which is
also $\cP\cT$ symmetric. Other nonlinear wave equations such as the generalized
KdV equation $u_t+u^ku_x+u_{xxx}=0$ and the Sine-Gordon equation $u_{tt}-u_{xx}+
g\sin u=0$ are $\cP\cT$ symmetric as well.

The striking observation that there are many nonlinear wave equations possessing
$\cP\cT$ symmetry suggests that one can generate many families of new complex
nonlinear $\cP\cT$-symmetric wave equations by following the same procedure that
was used in quantum mechanics [see (\ref{e1})]. One should then try to discover
which properties of the original wave equations are preserved.\footnote{An
alternative possibility for study is to examine inverse scattering problems and
isospectral flow using $\cP\cT$-symmetric potentials $u(x,t)$. We reserve this
research direction for a future paper.}

In this brief note we limit our discussion to the complex $\cP\cT$-symmetric
extension of the KdV equation:
\begin{equation}
u_t-iu(iu_x)^\epsilon+u_{xxx}=0,
\label{e3}
\end{equation}
where $\epsilon$ is a real parameter. We now examine the remarkable properties
of some members of this family of complex $\cP\cT$-symmetric equations. We will
emphasize the properties that members of this class have in common.

\vspace{.5cm}

\noindent
{\it Case} $\epsilon=1$: When $\epsilon=1$, (\ref{e3}) reduces to the KdV
equation (\ref{e2}). The KdV equation has an infinite number of conserved
quantities \cite{r9}. The first two are the momentum $P$,
\begin{equation}
\frac{d}{dt}P=0,\qquad P=\int dx\,u(x,t),
\label{e4}
\end{equation}
and the energy $E$,
\begin{equation}
\frac{d}{dt}E=0,\qquad E=\half\int dx\,[u(x,t)]^2.
\label{e5}
\end{equation}
The Cauchy initial-value problem for the KdV equation can be solved exactly
because the system is integrable, and it is solved by using the method of
inverse scattering \cite{r10}. 

\begin{figure}[th]\vspace{3.65in}
\includegraphics{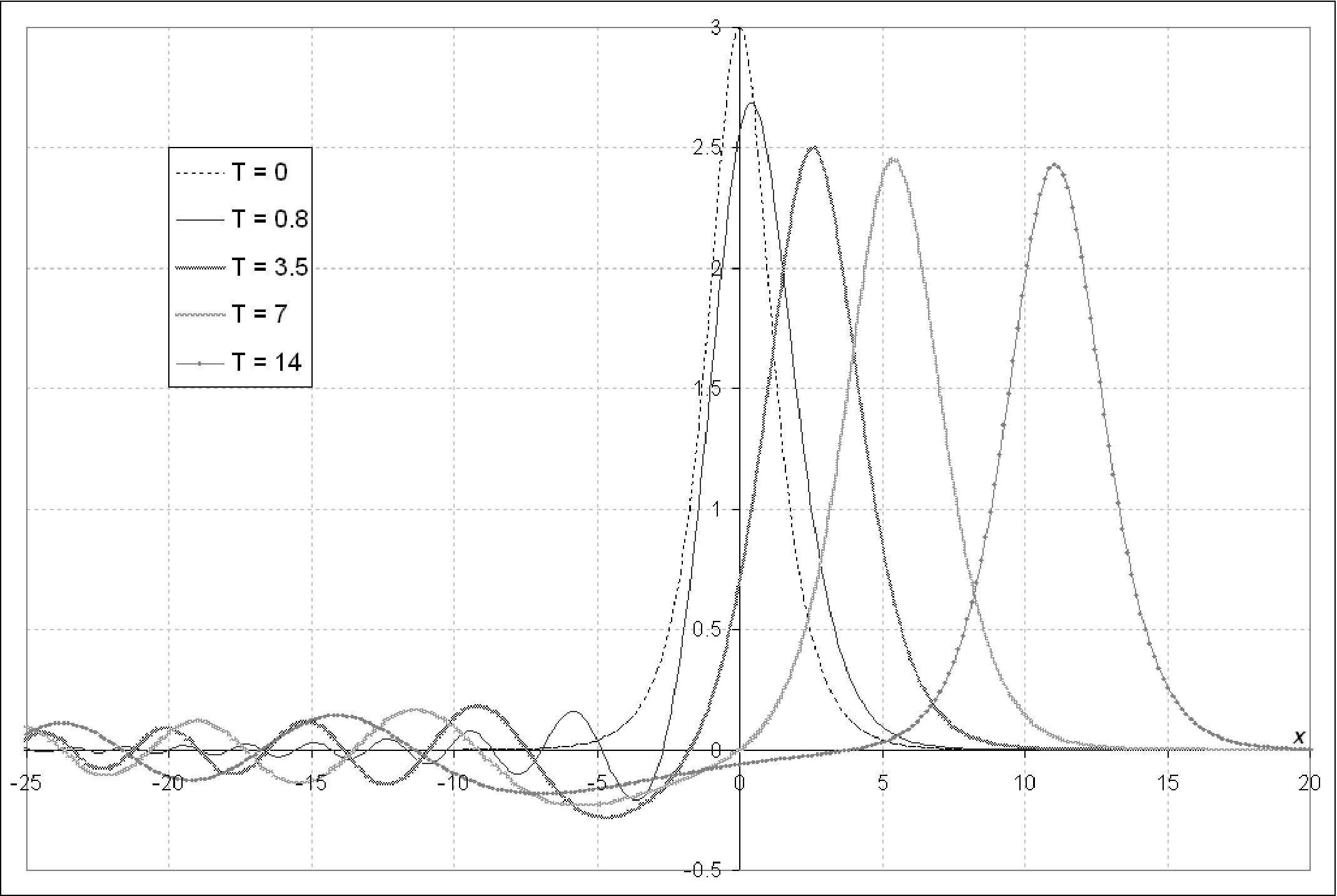}
\caption{Birth of a soliton. As the initial condition $u(x,0)=3\,{\rm sech}(x)$
evolves in time according to the KdV equation, a soliton of the form (\ref{e6})
moves to the right leaving a trail of residual radiation that propagates to the
left. The function $u(x,T)$ is plotted for $T=0$, $0.8$, $3.5$, $7$, and $14$.
\label{f1}}
\end{figure}

A solitary-wave solution to the KdV equation has the form
\begin{equation}
u(x,t)=3c\,{\rm sech}^2\left[\half\sqrt{c}\left(x-ct-x_0\right)\right],
\label{e6}
\end{equation}
where $c>0$ is the velocity. (In general, a {\it solitary-wave} solution $u(x,t)
=f(x-ct)$ to a partial differential equation is defined to be a wave that
propagates at constant velocity $c$ and whose shape does not change in time. In
this paper we also require that $f(z)\to0$ as $|z|\to\infty$.) These solitary
waves are called {\it solitons} because as they evolve according to the KdV
equation they retain their shape when they undergo collisions with other
solitary waves. One can observe numerically how a soliton emerges from a
pulse-like initial condition. For example, for the initial condition $u(x,0)=3\,
{\rm sech}(x)$, we see in Fig.~\ref{f1} that the pulse sheds a stream of
wave-like radiation that travels to the left and gives birth to a right-moving
soliton of the form in (\ref{e6}).

\vspace{.5cm}

\noindent
{\it Case} $\epsilon=0$: Setting $\epsilon=0$ in (\ref{e3}) gives the linear
equation
\begin{equation}
u_t-iu+u_{xxx}=0.
\label{e7}
\end{equation}
To solve the initial-value problem for this equation, we substitute $u(x,t)=e^{
it}v(x,t)$ and reduce it to $v_t+v_{xxx}=0$. We then perform a Fourier transform
to obtain the solution in the form of a convolution of the initial condition and
an inverse Fourier transform:
\begin{equation}
v(x,t)=v(x,0)\ast {\cal F}^{-1}\left(e^{ip^3t}\right). 
\label{e8}
\end{equation}
The inverse Fourier transform of the exponential of a cubic is an Airy
function \cite{r11}. Thus, the exact solution for $u(x,t)$ is
\begin{equation}
u(x,t)=e^{it}(3t)^{-1/3}\int_{-\infty}^\infty ds\,u(x-s,0){\rm Ai}\left[
(3t)^{-1/3}s\right].
\label{e9}
\end{equation}

\begin{figure}[th]\vspace{2.5in}
\includegraphics{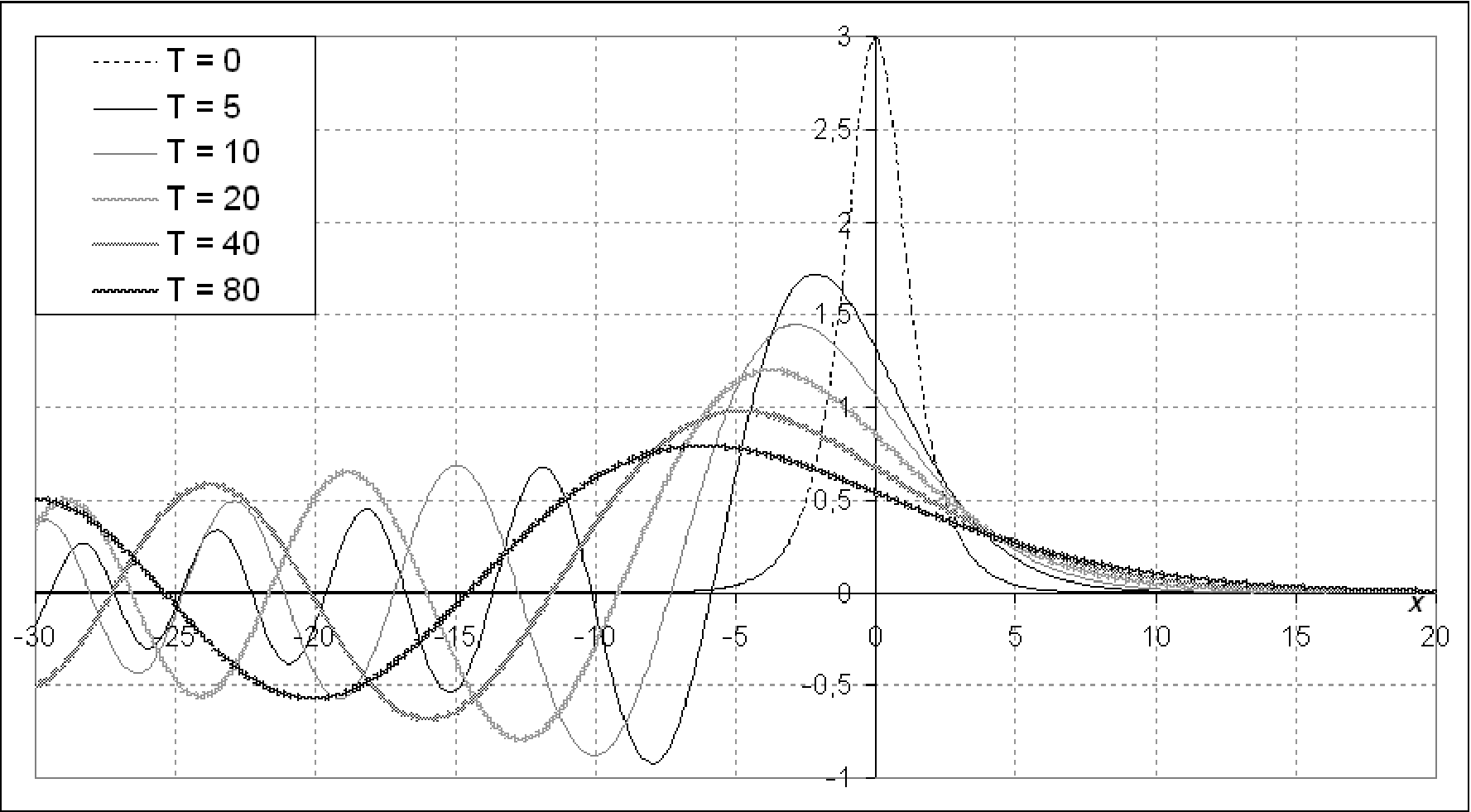}
\caption{Plot of $u(x,T)e^{-iT}$ as a function of $x$ for $T=0$, $5$, $10$,
$20$, $40$, and $80$, where $u(x,t)$ in (\ref{e9}) is calculated for the initial
condition $u(x,0)=3\,{\rm sech}(x)$. A comparison of this figure with 
Fig.~\ref{f1} shows that the solutions to (\ref{e3}) with $\epsilon=0$ and
$\epsilon=1$ are somewhat similar. However, in this case, while the initial
condition produces radiation that travels to the left, it does not give rise
to a soliton.
\label{f2}}
\end{figure}

The Airy function ${\rm Ai}(x)$ has a global maximum near $x=0$. For $x$ large
and positive ${\rm Ai}(x)$ decays exponentially, ${\rm Ai}(x)\sim(2\sqrt{\pi})^{
-1} x^{-1/4}\exp\left(-\frac{2}{3}x^{3/2}\right)$, and for $x$ large and
negative ${\rm Ai}(x)$ decays algebraically and oscillates, ${\rm Ai}(-x)\sim\pi
^{-1/2} x^{-1/4}\sin\left(\frac{2}{3}x^{3/2}+\frac{1}{4}\pi\right)$. Thus, the
qualitative behavior of ${\rm Ai}(x)$ resembles that in Fig.~\ref{f1}. If we
choose the initial condition $u(x,0)=3\,{\rm sech}(x)$ that was used to generate
Fig.~\ref{f1}, then apart from the phase $e^{it}$, we find that the solution
(\ref{e9}), which is shown in Fig.~\ref{f2}, resembles that for the KdV equation
in Fig.~\ref{f1} except that no soliton emerges from the initial condition.
There is only residual radiation that travels to the left. Thus we have extended
the KdV equation into the complex domain while preserving many of its
qualitative features.

\vspace{.5cm}

\noindent
{\it Case} $\epsilon=3$: If we set $\epsilon=3$ in (\ref{e3}), we obtain the
nonlinear wave equation
\begin{equation}
u_t-u(u_x)^3+u_{xxx}=0,
\label{e10}
\end{equation}
which has received only passing mention in the literature \cite{r12,r13}. The
only observations that have been made regarding this equation are that, apart
from translation invariance in $x$ and $t$, there is an obvious scaling solution
of the form $u(x,t)=\phi\left(xt^{-1/3}\right)$. Yet, this equation has an array
of rich and beautiful properties that have so far been overlooked. As we will
show, there are two conserved quantities, a momentum $P$ and an energy $E$,
which are the analogs of $P$ and $E$ in (\ref{e4}) and (\ref{e5}) for the KdV
equation. We will also show that there are traveling waves, and we will see how
a pulse-like initial condition gives birth to a traveling wave, just as in the
case of the KdV equation.

We derive the conserved quantities for the wave equation (\ref{e10}) in much
the same way that one finds the conserved quantities for the KdV equation
(\ref{e2}). However, the procedure is more elaborate. To find the momentum $P$,
we begin by integrating (\ref{e10}) with respect to $x$ and assume that $u(x,t)$
vanishes rapidly as $|x|\to\infty$. For the case of the KdV equation, this
procedure immediately gives the result in (\ref{e4}). However, for the equation
in (\ref{e10}) we have the result
\begin{equation}
\frac{d}{dt}\int dx\,u=\int dx\,u(u_x)^3.
\label{e11}
\end{equation}

Evidently, $\int u$ is not a conserved quantity because the right side of of
this equation does not vanish (in contrast to the KdV equation). To proceed we
introduce the identity
\begin{equation}
\int dx\,u^N(u_x)^3=\frac{2}{(N+1)(N+2)}\int dx\,u^{N+2}u_{xxx},
\label{e12}
\end{equation}
which is obtained by performing two integrations by parts. Using this identity
for the case $N=1$, we rewrite (\ref{e11}) as
\begin{equation}
\frac{d}{dt}\int dx\,u=\frac{1}{3}\int dx\,u^3u_{xxx}.
\label{e13}
\end{equation}
This equation suggests that we should multiply (\ref{e10}) by $u^3$, integrate
with respect to $x$, and use the identity (\ref{e12}) for the case $N=4$ to
obtain
\begin{equation}
\frac{1}{4}\,\frac{d}{dt}\int dx\,u^4=\frac{1}{15}\int dx\,u^6u_{xxx}
-\int dx\,u^3u_{xxx}.
\label{e14}
\end{equation}
We can combine (\ref{e13}) and (\ref{e14}) to eliminate the $\int u^3u_{xxx}$
term, but the right side will still not vanish because there will be a $\int u^6
u_{xxx}$ term.

Therefore, we must iterate this process by multiplying by $u^6$, $u^9$, $u^
{12}$, and so on, and then integrating with respect to $x$. We thus obtain the
following sequence of equations:
\begin{equation}
\frac{d}{dt}\int dx\,\frac{u^{3k+1}}{3k+1}=\int dx\,\frac{2u^{3k+3}u_{xxx}}
{(3k+2)(3k+3)}-\int dx\,u^{3k}u_{xxx},
\label{e15}
\end{equation}
where $k=0,1,2,3,\cdots$. We can now completely eliminate the right side if we
multiply the $k$th equation in (\ref{e15}) by
\begin{equation}
a_k=\frac{6^k\Gamma\left(k+\frac{1}{3}\right)}{(3k)!}A,
\label{e16}
\end{equation}
where $A$ is an arbitrary constant, and sum from $k=0$ to $\infty$. We conclude
that $\frac{d}{dt}P=0$, where the conserved quantity $P$ is given by
\begin{equation}
P=A\int dx\sum_{k=0}^\infty \frac{6^k\Gamma\left(k+\frac{1}{3}\right)u^{3k+1}}
{(3k+1)!}.
\label{e17}
\end{equation}

By a similar argument, we can construct a second conserved quantity $E$, $\frac{
d}{dt}E=0$, where $E$ is given by
\begin{equation}
E=B\int dx\sum_{k=0}^\infty \frac{6^k\Gamma\left(k+\frac{2}{3}\right)u^{3k+2}}
{(3k+2)!}
\label{e18}
\end{equation}
and $B$ is an arbitrary constant.

The summations in (\ref{e17}) and (\ref{e18}) can be performed in closed form
in terms of Airy functions, giving
\begin{eqnarray}
P&=&\int dx\,\int_0^{2^{1/3}u(x,t)} ds\,\left[{\rm Bi}(s)+\sqrt{3}{\rm Ai}(s)
\right],\nonumber\\
E&=&\int dx\,\int_0^{2^{1/3}u(x,t)} ds\,\left[{\rm Bi}(s)-\sqrt{3}{\rm Ai}(s)
\right],
\label{e19}
\end{eqnarray}
where we have taken $A=6^{1/3}/\pi$ and $B=6^{2/3}/\pi$. It is especially
noteworthy that the conserved quantity $E$ is strictly positive when $u(x,t)$ is
not identically 0, and thus it is reasonable to interpret $E$ as an energy. The
positivity property of the energy is maintained when $\epsilon$ changes from 1
(the KdV equation) to 3. We do not believe that (\ref{e10}) has more than two
conserved quantities.

Equation (\ref{e10}) is also similar to the KdV equation in that it has
solitary-wave solutions. To construct such a solution, we substitute $u(x,t)=f(x
-ct)$ into (\ref{e10}) to find the ordinary differential equation satisfied by 
$f(z)$:
\begin{equation}
-cf'(z)-f(z)[f'(z)]^3+f'''(z)=0.
\label{e20}
\end{equation}
It is only possible to solve this autonomous equation in implicit form. To do
so, we seek a solution of the form $f'(z)=G(f)$. The function $G$ satisfies
\begin{equation}
-2c-2fG^2(f)+\left[G^2(f)\right]''=0.
\label{e21}
\end{equation}
Making the further substitution $H(f)=G^2(f)$, we find that $H$ satisfies
\begin{equation}
H''(f)-2fH(f)=2c,
\label{e22}
\end{equation}
which is the inhomogeneous Airy equation, whose solution is expressed in terms
of the inhomogeneous Airy or Scorer function ${\rm Hi}$ \cite{r11}.

\begin{figure}[th]\vspace{2.0in}
\includegraphics{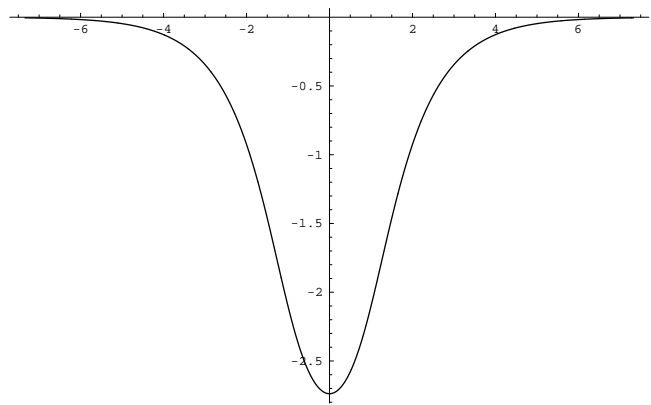}
\caption{Solitary-wave solution to the differential equation $u_t-u(u_x)^3+
u_{xxx}=0$. This negative-pulse solution has the form $u(x,t)=f(x-ct)$, where we
have taken $c=1$. The solitary wave is an even function of $z=x-t$ and it decays
like $e^{-|z|}$ for large $|z|$. Thus, it closely resembles the solitary-wave
solution in (\ref{e6}) for the KdV equation. At the negative peak the height of
this solitary wave is $-2.73802$.
\label{f3}}
\end{figure}

Unfortunately, because the solution to (\ref{e20}) is implicit, it is not easy
to determine immediately whether there are solitary-wave solutions [solutions
$f(z)$ that vanish as $|z|\to\infty$]. However, numerical analysis confirms that
there are indeed such solutions. In Fig.~\ref{f3} we have plotted the solitary
wave for $c=1$. Note that this wave is an even function of $z$ and it decays
like $e^{-|z|}$ for large $|z|$. Note also that the solitary-wave solution is a
{\it negative} pulse, rather than a positive pulse as with the KdV equation.

\begin{figure}[th]\vspace{3.1in}
\includegraphics{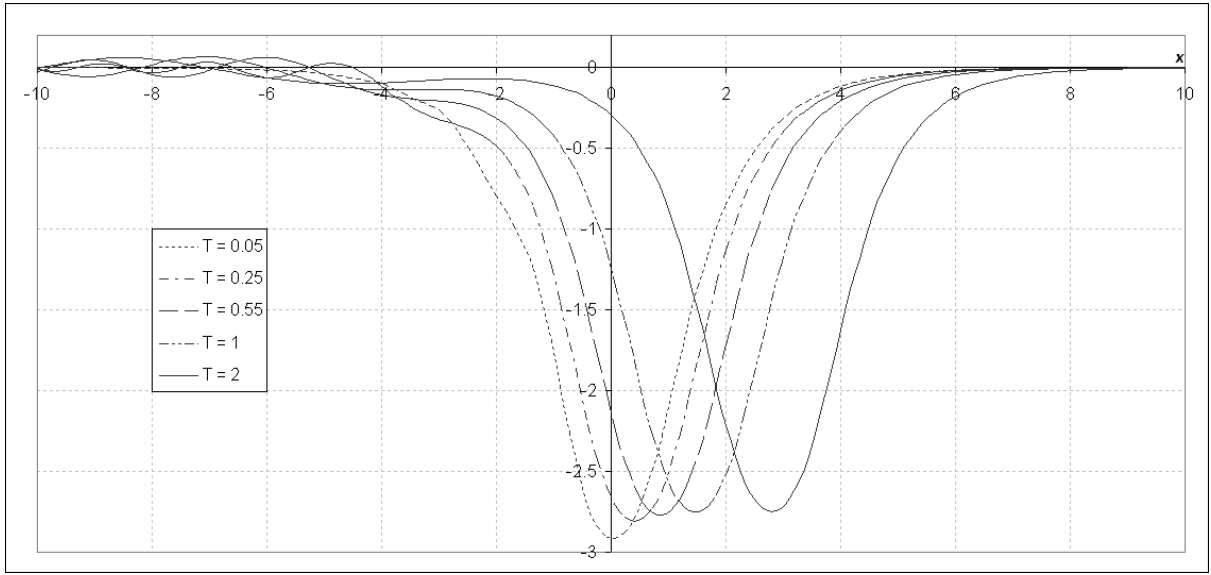}
\caption{Birth of a solitary wave. As the initial condition $u(x,0)=-3\,{\rm
sech}(x)$ evolves in time according to $u_t-u(u_x)^3+u_{xxx}=0$, a pulse that
approaches the shape of a solitary wave moves to the right leaving a trail of
residual radiation that propagates to the left. The wave $u(x,T)$ is shown for
the times $T=0.05,\,0.25,\,0.55,\,1,\,{\rm and}\,2$.
\label{f4}}
\end{figure}

As we saw in Fig.~\ref{f1} for the KdV equation, an initial pulse such as
$u(x,0)=-3\,{\rm sech}(x)$ for (\ref{e10}) gives birth to a solitary wave. As
shown in Fig.~\ref{f4}, this initial pulse emits radiation that travels to the
left and evolves into a right-going solitary wave (see Fig.~\ref{f3}). Computer
experiments suggest that these solitary waves are not solitons; that is, they
do not maintain their shape after a collision with another solitary wave.
Indeed, we would be surprised if (\ref{e10}) were an integrable system. The
quantum-mechanical Hamiltonian(\ref{e1}) ceases to be exactly solvable when
$\epsilon\neq0$, and in the same vein we expect that (\ref{e3}) is not
integrable when $\epsilon\neq0$.

\begin{figure}[th]\vspace{3.2in}
\includegraphics{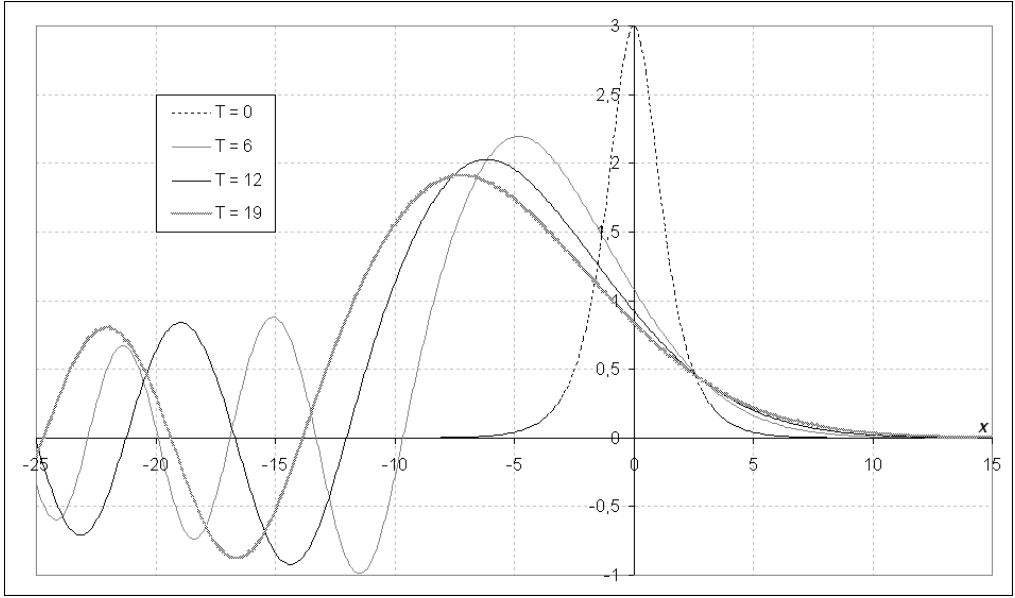}
\caption{The solution to $u_t-u(u_x)^3+u_{xxx}=0$ that evolves from the initial
condition $u(x,0)=3\,{\rm sech}(x)$. This initial condition generates radiation
that travels to the left, but it does not give rise to a solitary wave. The wave
$u(x,T)$ is shown for the times $T=0,\,6,\,12,\,{\rm and}\,19$.
\label{f5}}
\end{figure}

There are no positive solitary-wave solutions to (\ref{e10}). As we see in
Fig.~\ref{f5}, an initial pulse of the form $u(x,0)=3\,{\rm sech}(x)$ generates
a stream of radiation that travels to the left, but it does not give rise to a
solitary wave.

\begin{figure}[th]\vspace{2.3in}
\includegraphics{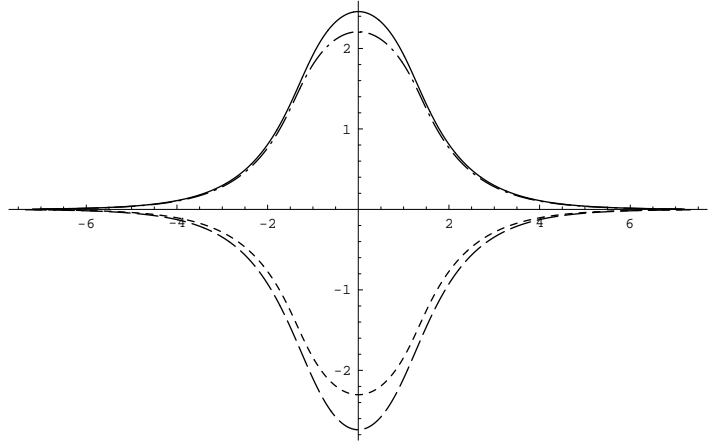}
\caption{Solitary-wave solutions of the form $f(x-t)$ to the differential
equation $u_t+(-1)^nu(u_x)^{2n+1}+u_{xxx}=0$ for $n=1,2,3,4$. Note that the
solutions are alternately positive and negative and gradually get wider as $n$
increases. At the stationary points the heights of the waves are $-2.73802$ ($n=
1$), $2.45839$ ($n=2$), $-2.30305$ ($n=3$), and $2.20797$ ($n=4$). The widths of
the waves at half-height are $3.15$, $3.14$, $3.19$, and $3.26$, respectively.
\label{f6}}
\end{figure}

\vspace{.2cm}
\noindent
{\it Case} $\epsilon=2n+1$: When $\epsilon=2n+1$ is an odd integer, the
nonlinear wave equation in (\ref{e3}) is real:
\begin{equation}
u_t+(-1)^nu(u_x)^{2n+1}+u_{xxx}=0.
\label{e23}
\end{equation}
For all values of $n$ there are solitary waves $u(x,t)=f(z)$, where $z=x-ct$,
and these waves are even functions of $z$. As $n$ increases, the solitary waves
alternate between being strictly positive and strictly negative functions and
gradually become wider (see Fig.~\ref{f6}).

In conclusion, we have shown how to extend the conventional KdV equation into
the complex domain while preserving $\cP\cT$ symmetry. The result is a large and
rich class of nonlinear wave equations that share many of the properties of the
KdV equation. In particular, we find that for some values of $\epsilon$ there
are conservation laws and solitary waves, and that arbitrary initial pulses can
evolve into solitary waves after they give off a stream of radiation. Airy
functions appear repeatedly in the analysis because the associated linear
equation that is satisfied when $u(x,t)$ is small is of Airy type.

In this paper we have just scratched the surface; one can begin with other $\cP
\cT$-symmetric nonlinear wave equations, such as the Camassa-Holm or the
generalized KdV equations, and study the properties of the resulting new complex
wave equations.

% QUESTION: --- WHAT IF TWO SOLITONS FOR $\epsilon=3$ COLLIDE??!!

\vspace{0.5cm}
\begin{footnotesize}
\noindent
We thank P.~Clarkson, D. Holm, H.~Jones, and B.~Muratori for useful discussions.
CMB is grateful to the Theoretical Physics Group at Imperial College, London,
for its hospitality. As an Ulam Scholar, CMB receives financial support from the
Center for Nonlinear Studies at the Los Alamos National Laboratory and he is
supported in part by a grant from the U.S. Department of Energy. DCB is
supported by The Royal Society.
\end{footnotesize}


\vspace{0.5cm}

\begin{thebibliography}{999}

\bibitem{r1} C.~M.~Bender and S.~Boettcher, Phys.~Rev.~Lett. {\bf 80}, 5243
(1998).

\bibitem{r2} C.~M.~Bender, S.~Boettcher, and P.~N.~Meisinger,
J.~Math.~Phys.~{\bf 40}, 2201 (1999).

\bibitem{r3} P.~Dorey, C.~Dunning and R.~Tateo, J.~Phys.~A: Math.~Gen.~{\bf 34},
L391 (2001); {\em ibid}. {\bf 34}, 5679 (2001).

\bibitem{r4} A.~Nanayakkara, Czech.~J.~Phys.~{\bf 54}, 101 (2004) and
J.~Phys.~A: Math.~Gen.~{\bf 37}, 4321 (2004).

\bibitem{r5} C.~M.~Bender, J.-H.~Chen, D.~W.~Darg, and K.~A.~Milton,
J.~Phys.~A: Math.~Gen.~{\bf 39}, 4219-4238 (2006).
% "Classical Trajectories for Complex Hamiltonians" 

\bibitem{r6} C.~M.~Bender and D.~W.~Darg (in preparation).

\bibitem{r7} C.~M.~Bender, D.~D.~Holm, and D.~W.~Hook, [arXiv: math-ph/0609068].

\bibitem{r8} R.~Camassa and D.~D.~Holm, Phys.~Rev.~Lett.~{\bf 71}, 1661 (1993).

\bibitem{r9} R.~M.~Miura, C.~S.~Gardner, and M.~D.~Kruskal, J.~Math.~Phys.~{\bf
9}, 1204 (1968).

\bibitem{r10} G.~B.Whitham, {\it Linear and Nonlinear Waves} (Wiley, New
York, 1974).

\bibitem{r11} M.~Abramowitz and I.~A.~Stegun, {\it Handbook of Mathematical
Functions} (Dover, New York, 1965), pp. 446-447.

\bibitem{r12} A.~D.~Polyanin and V.~F.~Zaitsev, {\it Handbook of Nonlinear
Partial Differential Equations} (Chapman \& Hall/CRC, Boca Raton, 2004),
pp.~526 and 529.

\bibitem{r13} W.~I.~Fushchych, N.~I.~Serov, and T.~K.~Ahmerov,
% On the conditional symmetry of the generalized KdV equation
Rep.~Ukr.~Acad.~Sci.~A {\bf 12}, 28 (1991).

\end{thebibliography}
\end{document}